# The Influence of Compressibility on the Restitution Coefficient for Viscoelastic Spheres in Low-Velocity Normal Impacts

Emanuel Willert*, e.willert@tu-berlin.de
Jean-Emmanuel Leroy, leroy@campus.tu-berlin.de
Maciej Satora, maciej.satora@campus.tu-berlin.de
Yannic Scholtyssek, scholtyssek@campus.tu-berlin.de

Technische Universität Berlin, 10623 Berlin, Germany

**Abstract.** The influence of compressibility on the coefficient of restitution in the normal impact of a rigid sphere onto a linear-viscoelastic compressible standard solid under quasi-static conditions is studied using a numerical solution procedure for the contact-impact problem based on the Method of Dimensionality Reduction. We find that the influence of compressible material behavior is most significant in parameter combinations with high energy dissipation in the bulk material during the impact. Thereby the restitution coefficient is usually underestimated, if the material is assumed to be incompressible. This misestimating can be very significant (relative errors of 100% and more are possible), even, if the actual Poisson ratio is very close to 0.5.

## 1 Introduction

Impacts of macroscopic particles determine the dynamics of granular gases [1], which has been applied, for example, in modelling the formation of Saturn's rings [2]. Impact testing is also a simple and fast option to analyze material behavior under dynamic loading [3].

Low-velocity impacts of viscoelastic bodies have been studied in numerous publications (see, for example [4]-[8] and the references therein) for different types of viscoelastic material behavior. In those works, however, almost exclusively (at least to the best our knowledge) incompressible media have been considered, which is mainly because of two reasons: first most biologically [9] or technologically [10] relevant viscoelastic materials can in good approximation be considered incompressible (the bulk modulus is usually several orders of magnitude higher than the shear modulus). On the other hand, accounting for compressibility and thus a stress response to a unit volume strain severely complicates the problem solution based on rigorous contact mechanics [11]. Nonetheless, there are, of course, compressible viscoelastic materials, for example the Earth's mantle [12], and the influence of compressibility in viscoelastic impacts represents an interesting and hitherto unsolved problem.

Very recently one of the authors ([13], [14]) has shown, how, based on the elastic-viscoelastic correspondence principle [15], the frictionless normal contact problem of compressible viscoelastic materials can be exactly reduced to the contact between a rigid indenter and an incompressible viscoelastic half-space with a properly defined shear relaxation (or, equivalently, shear creep) function. The methods, developed to study incompressible viscoelastic impacts, can, thus, be also applied to the analysis of compressible material behavior, if the appropriate (equivalent) incompressible shear relaxation function is used.

In the past years with the Method of Dimensionality Reduction (MDR) a very efficient and powerful tool has been developed to simulate and analyze axisymmetric viscoelastic contact problems for arbitrary material rheology and arbitrary loading histories (see [16], [17] and the references therein). The MDR has been proven [18] to exactly reproduce the contact solutions by Lee & Radok [19] for the compression phase and by Graham [20] and Ting [21] for the restitution phase.

Hence, in the present manuscript we will study the influence of compressibility on the restitution in viscoelastic impacts based on a numerical procedure within the framework of MDR.

*corresponding author

## 2 Problem formulation

For simplicity we will consider the frictionless normal impact of a rigid sphere with mass $m$ and radius $R$ and initial velocity $v_0$ onto a homogeneous, isotropic linear-viscoelastic half space with the shear relaxation function

$$G(t) = G_\infty + G_1 \exp\left(-\frac{G_1}{\eta} t\right). \tag{1}$$

Here $G_\infty$ is the static modulus, $G_0 = G_\infty + G_1$ the (instantaneous) glass modulus and $\eta$ the shear viscosity. The material with the relaxation function (1) is often referred to as the "three-element standard solid", which exhibits most qualitative features of "real" elastomers, like rubber. The behavior of the viscoelastic medium under hydrostatic stress conditions shall be completely elastic, i.e. the bulk relaxation function is given by

$$K(t) \equiv K_0 = const, \tag{2}$$

with the modulus $K_0$. To characterize compressibility we use the static Poisson ratio

$$\nu := \frac{3K_0 - 2G_\infty}{6K_0 + 2G_\infty} \in [-1; 0.5[. \tag{3}$$

A scheme of the impact problem is shown in Fig. 1. Note, that the more general problem of two colliding viscoelastic spheres exhibits no qualitatively different features.

As we want to gain a qualitative understanding of the influence of compressibility, we will further assume, that the impact is quasi-static, i.e. the impact velocity is much smaller than the smallest speed of wave propagation in the viscoelastic medium, and that all deformations remain small. We stress, that both assumptions, admittedly, pose severe restrictions in the case of very soft materials. However, accounting for wave propagation and large deformations, i.e. geometric and constitutive nonlinearities, drastically complicates the problem and will probably give little additional insight into the influence of compressiblity.

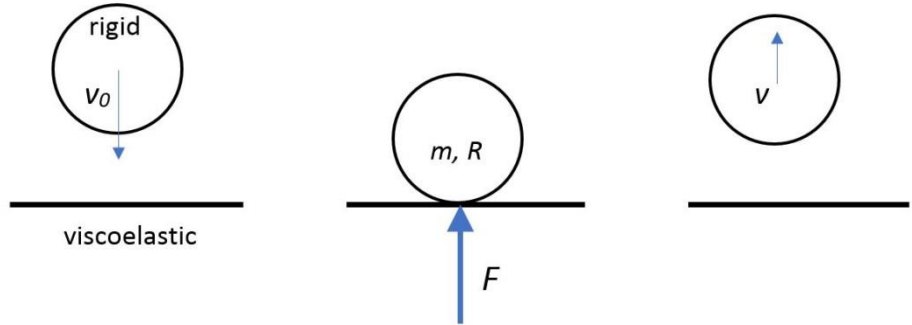

**Fig. 1** Scheme of the studied quasi-static normal impact problem.

## 3 Numerical Solution Procedure based on the MDR

Within the MDR the three-dimensional axisymmetric contact between an indenter with the profile $f(r)$ and a viscoelastic half-space is mapped onto the contact of an equivalent plain profile [16]

$$g(x) = |x| \int_0^{|x|} \frac{f'(r) dr}{\sqrt{x^2 - r^2}} \tag{4}$$

with a one-dimensional foundation of independent linear viscoelastic elements. The spherical profile can, in the vicinity of the contact, be approximated by the parabola

$$f(r) = \frac{r^2}{2R}, \tag{5}$$

which will readily give the equivalent profile

$$g(x) = \frac{x^2}{R}. \qquad (6)$$

As shown in [13], to capture the material behavior given by Eqs. (1) and (2), the elements of the viscoelastic foundation are given by the rheological model depicted in Fig. 2.

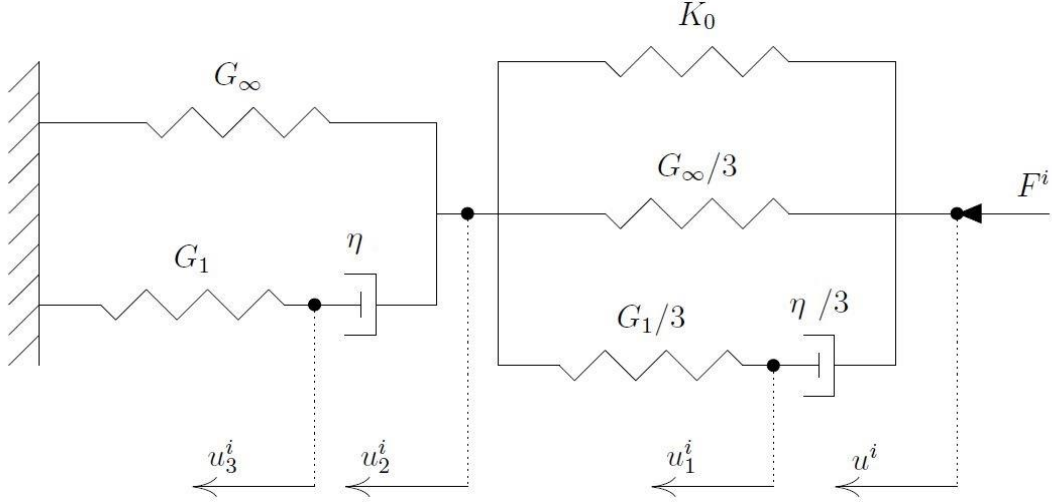

**Fig. 2** Rheological model of the incompressible viscoelastic material, which in frictionless normal contact problems is equivalent to the compressible standard solid defined by Eqs. (1) and (2).

The displacement of the outer degree of freedom $u(x,t)$ for each element in contact is enforced by the impacting sphere,

$$u(x,t) = d(t) - g(x), \quad |x| \leq a(t), \qquad (7)$$

with the indentation depth $d$ and the contact radius $a$. During the compression stage of the impact, all elements come into contact geometrically. Due to the fact, that we only consider quasi-static behavior, we can formulate the equilibrium conditions for the inner and outer degrees of freedom of an element $i$ at the position $x_i = i\Delta x$, with the element spacing $\Delta x$ [17]:

$$\begin{aligned} 0 &= \eta(\dot{u}_3^i - \dot{u}_2^i) + G_1 u_3^i, \\ 0 &= \frac{\eta}{3}(\dot{u}_1^i - \dot{u}^i) + \frac{G_1}{3}(u_1^i - u_2^i), \\ 0 &= G_1 u_3^i + G_\infty u_2^i + \left(\frac{G_\infty}{3} + K_0\right)(u_2^i - u^i) + \frac{G_1}{3}(u_2^i - u_1^i), \\ F^i &= 4\Delta x \left(G_\infty u_2^i + G_1 u_3^i\right). \end{aligned} \qquad (8)$$

If the element force $F^i$ would be negative, it is set to zero and the respective element leaves contact. The total contact force is given by the sum over all element forces, so the equation of motion reads

$$m\ddot{d} = -\sum_i F^i. \qquad (9)$$

The impact is over, if all elements have left contact. Eqs. (7) to (9) give a closed system of ordinary differential equations, which can be solved by any time-integration scheme. We chose an explicit Euler method with time step $\Delta t$, because the computational operations are fast and simple and the time step can therefore be set small enough to avoid stability problems.

Note, that, during restitution, it is not necessary to trace the displacements of the elements, which have left contact, because contact can never be re-established anywhere.

## 4 Results

As can be shown by dimensional analysis and numerical experiments, the restitution coefficient (CoR)

$$e := -\frac{v}{v_0}, \qquad (10)$$

with the rebound velocity $v$, for the stated problem only depends on three dimensionless parameters

$$p_1 = \nu, \quad p_2 = \frac{G_\infty}{G_1}, \quad p_3 = \eta \left( \frac{v_0 R}{m^2 G_\infty^3} \right)^{1/5}. \qquad (11)$$

The parameter $p_3$ is roughly related to the ratio between the impact duration and the characteristic relaxation time of the elastomer. Fig. 3 and Fig. 4 show contour diagrams of the CoR as a function of the parameters $p_2$ and $p_3$ in logarithmic scale for $\nu = 0.45$ and for the incompressible limit, which has been described in detail in [7]. It appears that compressibility mainly influences the CoR in regions, where the latter one is small, i.e. for parameter combinations with high dissipation.

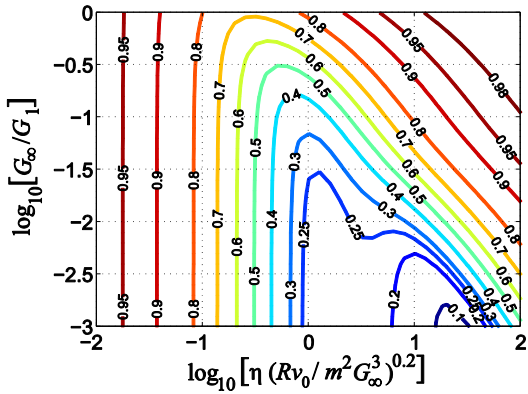

**Fig. 3** Contour isoline diagram of the CoR for the impact of a rigid sphere onto a compressible standard solid as a function of the remaining parameters for $\nu = 0.45$. All free input parameters have been generated randomly.

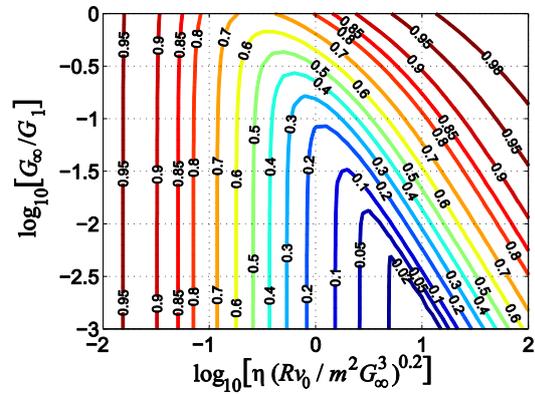

**Fig. 4** Contour isoline diagram of the CoR for the impact of a rigid sphere onto an incompressible standard solid as a function of the remaining parameters. All free input parameters have been generated randomly.

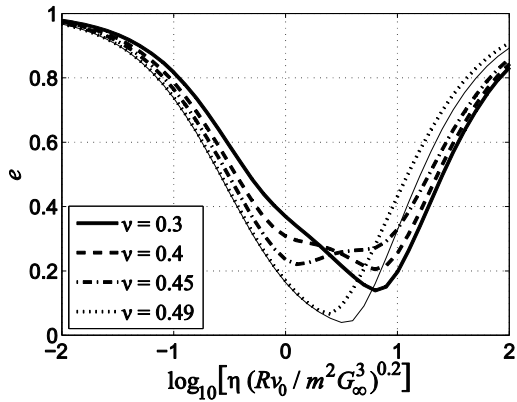

**Fig. 5** Coefficient of restitution $e$ for the impact of a rigid sphere onto a compressible standard solid for different values of the static Poisson ratio and $\log_{10}(G_\infty/G_1) = -2$. Thin line denotes the incompressible limit. All free input parameters have been generated randomly.

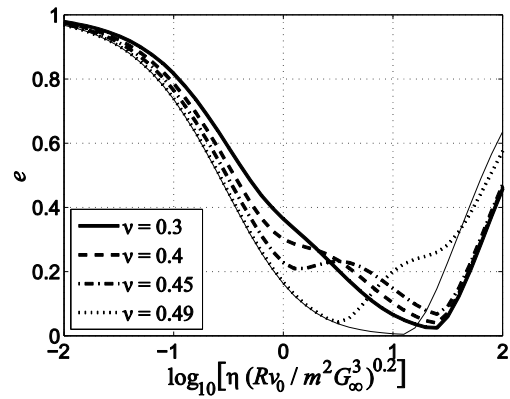

**Fig. 6** Coefficient of restitution $e$ for the impact of a rigid sphere onto a compressible standard solid for different values of the static Poisson ratio and $\log_{10}(G_\infty/G_1) = -3$. Thin line denotes the incompressible limit. All free input parameters have been generated randomly.

This is demonstrated again in Fig. 5 and Fig. 6, where the CoR is shown as a function of the parameter $p_3$ for different values of the static Poisson ratio $\nu$ and the ratio of shear moduli $p_2$. Obviously the difference between the incompressible and the compressible solution is high, if the CoR itself is small. This is the case, if $p_3$ is of the order of unity, i.e. if the impact has the same time scale as the relaxation in the viscoelastic material (note, that, if the material has several relaxation times, the faster relaxation processes will inhibit the return to the elastic domain for fast impact velocities in the right branches of the upper figures). This can be explained the following way: in the described parameter areas, the stiffness of the viscoelastic material is highly important for the amount of dissipation; as the incompressible material is stiffer than the respective compressible one, the CoR is altered quite significantly (compare, for example, the curve for ν = 0.49 with the incompressible limit around the value $p_3 = 10$ in Fig. 6).

# 5  Discussion

We studied the influence of compressibility on the coefficient of restitution in the normal impact of a rigid sphere onto a linear-viscoelastic compressible standard solid under quasi-static conditions using a numerical solution procedure for the contact-impact problem based on the MDR.

We find that the influence of compressible material behavior is most significant in parameter combinations with high energy dissipation in the bulk material during the impact. Thereby the restitution coefficient is usually underestimated, if the material is assumed to be incompressible. This misestimating can be very significant (relative errors of 100% and more are possible), even, if the actual Poisson ratio is very close to 0.5.

The stated assumptions (linear viscoelastic material with only one relaxation time, quasi-static conditions, small deformations) will, of course, probably be broken in practical applications. However, our main qualitative conclusion, that compressibility can have a high influence on the restitution, if the CoR is already small, should, nonetheless remain valid.